\documentclass[iop,apj]{emulateapj}

\shorttitle{Mass Models of Dwarf Spheroidal Galaxies}

\shortauthors{Mashchenko}

\begin{document}

\title{Mass Models of Dwarf Spheroidal Galaxies with Variable Stellar Anisotropy. I. Jeans Analysis}

\author{Sergey Mashchenko}

\affil{Department of Physics and Astronomy, McMaster University,
Hamilton, ON, L8S 4M1, Canada; syam@physics.mcmaster.ca}

\begin{abstract}
  Using a flexible galactic model with variable stellar velocity
  anisotropy, I apply the classical Jeans mass modeling approach to the
  five dwarf spheroidal galaxies with the largest homogeneous datasets
  of stellar line-of-sight velocities (between 330 and 2500 stars per
  galaxy) -- Carina, Fornax, Leo~I, Sculptor, and Sextans. I carry out
  an exhaustive model parameter search, assigning absolute
  probabilities to each parameter combination. My main finding is that
  there is a well defined radius (unique for each galaxy) where the
  Jeans analysis constraints on the enclosed mass are tightest, and
  are much better than the constraints at previously suggested radii
  (e.g. 300~pc). For Carina, Fornax, Leo~I, Sculptor, and Sextans the
  enclosed DM mass is $0.94 \pm 0.20$ (at 410~pc), $7.1 \pm 0.9$ (at
  925~pc), $1.75 \pm 0.20$ (at 390~pc), $2.59 \pm 0.42$ (at 435~pc),
  and $2.3 \pm 0.9$ (at 1035~pc), respectively (two-sigma
  uncertainties; in $10^7$~M$_\odot$ units). Local DM density has the
  tightest constraints at smaller (and also unique for each galaxy)
  radii. The largest central DM density constraint is for Sculptor:
  $\rho_0\gtrsim 0.09$~$M_\odot$~pc$^{-3}$ (at two-sigma level). I
  show that the DM density logarithmic slope is totally
  unconstrained by the Jeans analysis at all the radii probed by the data
  (and not just at the center, as was demonstrated before). Stellar
  velocity anisotropy has only very weak constraints. In particular,
  pure central tangential anisotropy is ruled out at better than two
  sigma level for three dwarfs, and the data are
  consistent with the global stellar velocity isotropy for all the
  five galaxies. 
\end{abstract}


\section{INTRODUCTION}

Dwarf spheroidals (dSphs) are very faint galaxies (some of the recently
discovered dSph satellites of Milky Way are the faintest known
galaxies, \citealt{mar07}) which tend to cluster around large galaxies
\citep{mat98}. Among the 50 or so Local Group dSphs discovered so far,
only two (Tucana and Cetus) are relatively isolated; the rest are
located in the vicinity of either Milky Way or M31, and are believed
to be their satellites. Despite their dull appearances (no ongoing
star formation; little or no interstellar medium; \citealt{mat98}),
dSphs are fascinating objects -- largely due to the fact that they
contain significant quantities of dark matter (DM), and as such
represent the smallest observed scale of DM clustering.

State of the art cosmological simulations predict that large galaxies
should contain many smaller sub-halos which have not had time to be
completely destroyed by the tidal field of the host halo. dSphs are
believed to represent at least some of this substructure
\citep{moo99}.

dSphs present two major challenges to the standard ($\Lambda$CDM)
cosmological model. First, despite the recent advances on both
theoretical and observational sides, there still appears to be a
factor of a few discrepancy between the observed and predicted
numbers of Milky Way satellites -- so called ``missing satellites
problem'' \citep{moo99}. Second, some indirect evidence (such as the
``timing paradox'' for the five globular clusters of Fornax dSphs;
\citealt{goe06}) suggests that DM distribution in dSphs has a flat
core, which is in line with what is deduced for larger dwarf and low
surface brightness galaxies \citep{vdb01,deb02,mar02,gen05}, but is at
odds with the results of many cosmological simulations, predicting
central divergent DM cusps with the logarithmic slope $\gamma\sim 1$
\citep{NFW}.  Different theoretical mechanisms suggested to rectify
this ``cusp--core'' problem produce different velocity anisotropy
profiles for stars. For example, our stellar feedback mechanism
\citep{mas06b,mas08}, where the DM cusp is flattened via gravitational
heating from the interstellar gas sloshed around by supernovae,
results in isotropic stellar cores.  On the other hand, the dynamical
friction mechanism of \citet{elz01}, where the cusp is heated by
massive baryonic clumps passively spiraling in towards the galactic
center, should produce noticeable central tangential anisotropy
\citep{ton06}. Constraining stellar anisotropy profiles in spheroidal galaxies
can hence be a valuable indirect way of identifying the
correct theoretical mechanism of DM core formation.

\citet{aar83} was the first who deduced the presence of significant
amounts of DM in a dSph (Draco), based on the measurements of the
line-of-sight velocities of only three stars. Since then, the number of stars
in dSphs with known line-of-sight velocities has grown dramatically:
recently published homogeneous datasets have up 2500 stars per galaxy
(for Fornax; \citealt{wal09c}). Mass modeling of dSphs has
dramatically improved as well: from simple global estimates based on
the virial theorem \citep{aar83}, to spatial models with strong
simplifying assumptions (such as ``mass follows light'';
\citealt{mat98}), to more recent modeling efforts with fewer
assumptions \citep{wal09c,wol09,str08}.

The classical method of dSph mass modeling is to solve
the spherical Jeans equation for a range of models with different
DM density profiles, and then to compare the predicted stellar line-of-sight
velocity dispersion profiles with the observed ones -- using either $\chi^2$
or maximum likelihood techniques to find the best models. The
spherical symmetry assumption is reasonably accurate given the
spheroidal appearances of dSphs (ellipticity $e$ is usually $\sim 0.3$
or less), absence of disk-like structures, and negligible
rotation in these systems \citep{mat98}. To solve the spherical Jeans
equation (that is, to derive the stellar velocity dispersion as a
function of radius), one has to specify two functions: total (DM $+$
stars) density, and stellar velocity anisotropy ($\beta$) profiles. To
reduce the dimensionality of the problem, one usually assumes
a certain shape of the $\beta$ radial profile -- either a constant, 
or the Osipkov-Merritt profile (purely isotropic at the center, purely radially
anisotropic in the infinity; \citealt{osi79,mer85}). 

There are no reasons to believe that stellar anisotropy is
constant in dSphs. For example, simple dynamical models which start with an
initially non-equilibrium stellar configuration inside the DM halo (``cold
collapse'', \citealt{mas04}, and ``hot collapse'', \citealt{mas05}) tend
to produce the stellar velocity isotropy at the center and a variable
degree of radial anisotropy in the outskirts of the relaxed stellar
body. As already mentioned, the central parts of dwarf galaxies can be
significantly affected by dense and violent baryons, which may either
randomize stellar velocities \citep{mas06b,mas08} or induce a
significant tangential anisotropy \citep{ton06}. On the observational
side, dSphs are known to be non-homogeneous objects, with younger,
more metal-rich and kinematically colder star populations concentrated
towards the center \citep{tol04}; it would be strange if the stellar
velocity anisotropy would be the same across these different
populations of stars. 

Given the expectation that $\beta$ is not constant across a dSph and
that the important properties of dSphs (such as the central
logarithmic slope of the DM density) are  strongly
degenerate with respect to the unknown stellar velocity anisotropy
\citep{str08,wal09c}, proper mass modeling of dSphs has to include
radially variable stellar anisotropy. To the best of my knowledge, the
only example of a full-fledged Jeans mass modeling of dSphs with
variable $\beta$ is the work by \citet{str08}. Here I outline the main
differences of my approach with that of \citet{str08}.

\begin{enumerate}

\item In this paper, I carry out mass modeling of the limited number
  (five; Carina, Fornax, Leo~I, Sculptor, and Sextans) of Galactic
  dSphs with the highest quality, homogeneous data: large ($> 300$
  stars) homogeneous datasets of stellar line-of-sight velocities
  \citep{wal09c} and accurate stellar surface brightness profiles
  derived in a homogeneous manner in this paper. This is in contrast
  to the work of \citet{str08}, who analyzed a much larger set of
  Local Group dSphs, with heterogeneously derived data and a wide
  range of data quality.

\item I employ a ``brute force'' optimization while searching for the best
  fitting models. This approach generated a wealth of statistical
  data, which can be used for testing a variety of different
  hypotheses about the distribution of stars and DM in dSphs. The data
  are available online.

\item The ``brute force'' approach coupled with a flexible dSph model
  allowed me to find the dSph parameters which the Jeans analysis can
  constrain very well. Specifically, I found that the total enclosed
  mass in a dSph can be constrained to a high accuracy (two-sigma
  uncertainty as good as $\pm $15\%, for Fornax) at a certain radius, which is
  different for different dwarfs. This information can be valuable for
  matching the Galactic dSphs to the predictions of cosmological
  simulations. Similarly, I derived tight constraints on the local DM
  density at a certain radius (unique for each galaxy), which can be
  used in the research aimed at detecting DM in dSphs via its
  annihilation signal.

\item Unlike \citet{str08}, I properly account for the self-gravity
  of stars.  This can be important for dSphs with dense stellar cores
  (e.g. Sculptor).

\item I use an advanced (with the variance due to random locations of
  stars within one radial bin removed; see \S~\ref{good}) $\chi^2$
  model fitting approach. Unlike \citet{str08}, who employed a
  maximum likelihood technique, my approach is insensitive to the
  (unknown) shape of the line-of-sight star velocity probability
  distribution function (PDF).

\end{enumerate}

I had two main goals for this paper. First, I wanted to find out which
(if any) dSph parameters can be meaningfully constrained via the
classical, Jeans mass modeling approach, if such an analysis is pushed
to the extreme (large homogeneous observational datasets of stellar
line-of-sight velocities; a very flexible galactic model; careful
numerical analysis consuming a significant -- $3\times 10^5$~cpu hours
in my case -- amount of supercomputing time; a ``brute force''
optimization which ensures that no hidden ``valleys'' and local and
global minima in the multi-dimensional likelihood manifold are
missed). The second goal was to generate high-quality data which can
be used to dramatically reduce computational time (by restricting the
model parameter space to be explored) in future post-Jeans (analyzing
the full shape of the stellar velocity PDF) mass modeling efforts --
which will be the subject of Paper~II in this series.

As the dynamical state of the outer parts of dSphs is still a matter
of debate (specifically, some authors, e.g. \citealt{mun06}, argue
that the observed outskirts of these galaxies are undergoing a severe
tidal disruption and hence are not in dynamic equilibrium, which would
invalidate the Jeans analysis for those parts; but see \citealt{seg07}
for the opposite example), in the present study I put the main
emphasis on recovering the {\it inner} structure of dSphs.
Unfortunately, the most interesting aspect of the inner structure of
these galaxies -- the ``DM cusp or core'' question -- cannot be
addressed via the Jeans analysis alone, as the central logarithmic
slope of the DM density profile $\gamma$ is known to be degenerate in
this method.  This follows from the properties of the spherical Jeans
equation \citep {BT}, and is clearly visible even in the mass modeling
with a constant stellar anisotropy \citep{wal09c}. In this paper I obtain a more general
result that $\gamma(r)$ is completely unconstrained by the Jeans
analysis not only at the center ($r=0$), but also at any other radius
$r$ within the stellar body of the dwarf.

This paper is organized as follows. Section~\S~\ref{model} describes
the dSph model. Section~\S~\ref{obs_data} discusses the observational
data (stellar line-of-sight velocities and stellar surface brightness
profiles).  Section~\S~\ref{fitting} gives a detailed description of my
$\chi^2$ optimization technique. Section~\S~\ref{results} presents the
main results of this study. The conclusions are presented
in Section~\S~\ref{conclusions}.

\section{MODEL}
\label{model}

Assuming a spherical symmetry, the Jeans equation for a two-component
(DM $+$ stars) system can be written as

\begin{equation}
\label{Jeans}
\frac{1}{\rho_*}\frac{d(\rho_* \sigma_r^2)}{dr} + \frac{4\eta}{1+\eta}\frac{\sigma_r^2}{r}
= -\frac{d\Phi}{dr}.
\end{equation}

\noindent Here $r$ is the distance from the center,  $\rho_*$, $\sigma_r$, and $\eta$ are the stellar density,
radial velocity dispersion, and anisotropy, and $\Phi$ is the total
(DM $+$ stars) gravitational potential. We introduced the anisotropy
parameter $\eta$ in \citet{mas06a}. It is defined as
$\eta=(\sigma_r^2-\sigma_t^2)/(\sigma_r^2+\sigma_t^2)$, and is related
to the more conventional anisotropy parameter $\beta$ through
$\eta=\beta/(2-\beta)$ and $\beta=2\eta/(1+\eta)$. (Here $\sigma_t$ is
the one-dimensional tangential velocity dispersion.) Unlike $\beta$,
the parameter $\eta$ is conveniently symmetric: it is equal to $-1$, 0, and 1 for
purely tangential, isotropic, and purely radial orbits, respectively.
(The corresponding $\beta$ values are $-\infty$, 0, and 1, respectively.)

I assume that the stellar anisotropy $\eta$ smoothly varies between the two
asymptotic values, $\eta_0$ at the center and $\eta_1$ in the infinity:

\begin{equation}
\label{eta}
\eta(r) = \frac{\eta_0 A+\eta_1 (r/r_a)^2}{A+(r/r_a)^2}
\end{equation}

\noindent \citep{bae07,str08}. Here $r_a$ is the anisotropy transition 
radius, and $A\equiv (1+\eta_1)/(1+\eta_0)$.

The DM halo is assumed to have the following density profile:

\begin{equation}
\label{eta}
\rho(r)=\frac{\rho_s}{(r/r_s)^\gamma (1+r/r_s)^{3-\gamma}}.
\end{equation}

\noindent (Here $r_s$ and $\rho_s$ are the DM scaling radius and
density, respectively.) This is a generalization of the \citet*[NFW]{NFW}
profile, with an arbitrary central logarithmic density slope $\gamma$
and the outer slope of $3$. In the limit $\gamma=1$ we recover the
theoretical NFW profile; when $\gamma=0$, we obtain a flat-cored
profile, which is almost identical to the observationally inferred
\citet{bur95} profile. Stars are assumed to have a generalized Plummer
density profile (see eq.~(\ref{rho_*}) in \S\ref{SB}). The DM
contribution to the total gravitational potential $\Phi$ is obtained
by numerical integration of the corresponding Poisson equation; the
stellar contribution has simple analytical solutions (different for
different values of the stellar outer density slope $\alpha$).

To compare the solution of the Jeans eq.~(\ref{Jeans}) with
observations, one has to compute the stellar line-of-sight velocity
dispersion, $\sigma_{\rm los}$, as a function of the projected radius,
$R$:

\begin{equation}
\label{sigma_los}
\sigma_{\rm los}^2(R)=\frac{2}{\Sigma(R)}\int\limits_R^\infty\left(1-\frac{2\eta R^2}{(1+\eta) r^2}\right)
\frac{\rho_*\sigma_r^2 r\: {\rm d} r}{(r^2-R^2)^{1/2}}.
\end{equation}

\noindent \citep[p.~208]{BT}  Here $\Sigma(R)$ is the stellar surface density.

Overall, the model has six free parameters: $r_s$, $\rho_s$, $\gamma$
(characterize the DM distribution), and $\eta_0$, $\eta_1$, $r_a$ (characterize
the stellar anisotropy profile).

\section{OBSERVATIONAL DATA}
\label{obs_data}

\subsection{Surface brightness profiles}
\label{SB}

For Jeans mass modeling it is essential
to know stellar density profiles of dSphs as accurately as
possible. These can be derived from surface brightness profiles if one
assumes a spherical symmetry and a certain value for the stellar
mass-to-light ratio.  As Galactic dSphs are located within 300~kpc
from the Sun and are fully resolved into stars, one has to employ star
count techniques to derive their surface brightness profiles.

Traditionally, a few simple analytical stellar surface density
profiles have been used for mass modeling of dSphs: empirical
\citet{kin62}, Plummer \citep{BT}, and \citet{ser63}. We showed in
\citet{mas06a} that the surface density profile of the form

\begin{equation}
\label{Sigma}
\Sigma=\Sigma_0 \left[ 1 + (R/b)^2\right]^{-(\alpha-1)/2},
\end{equation}

\noindent corresponding to the ``generalized Plummer law'',

\begin{equation}
\label{rho_*}
\rho_*=\rho_{*,0} \left[ 1 + (r/b)^2\right]^{-\alpha/2},
\end{equation}

\noindent describes the surface brightness profile of Draco dwarf 
spheroidal extremely well. (Here $R$ and $r$ are the projected and spatial
distances from the dwarf's center, $\Sigma_0$ and $\rho_{*,0}$ are the
central stellar surface and volume densities, $b$ is the ``core
radius'', and $\alpha$ is an integer which should be $\geqslant 4$ for
finite total stellar mass models.) The classical Plummer law is recovered 
when $\alpha=4$.

\begin{figure*}
\epsscale{0.8}
\plotone{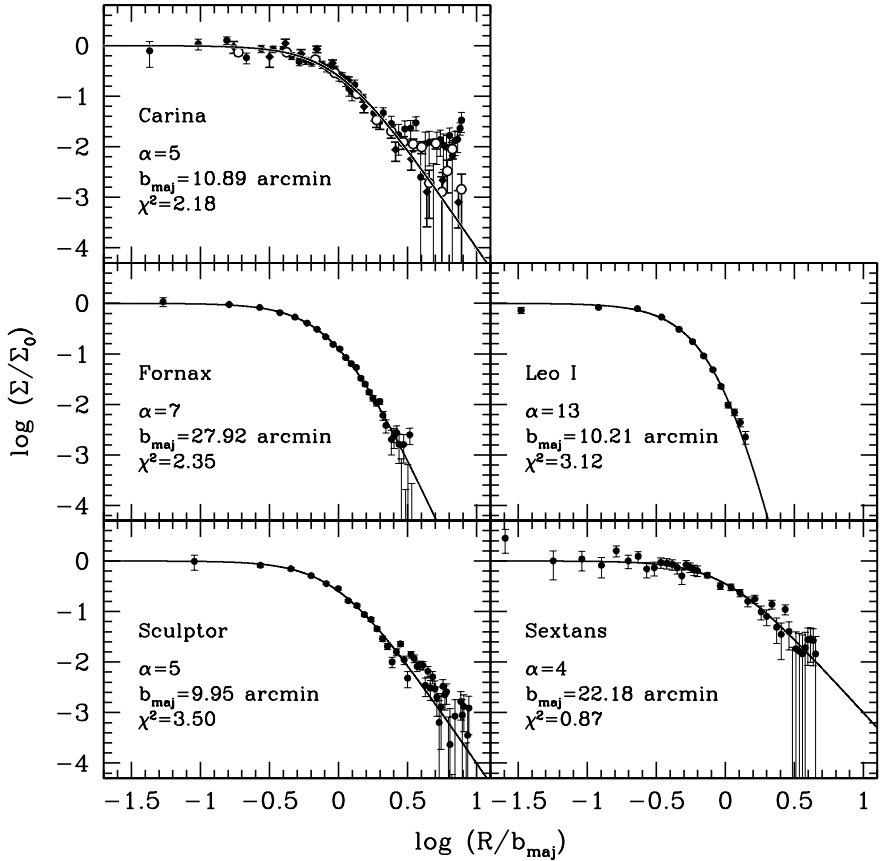}
\caption {Normalized radial surface brightness profiles.
Points with error bars show the observed star counts; the lines show the
best fitting generalized Plummer models. For Carina, the data from
\citet{IH95}, \citet{maj00}, and \citet{mun06} are shown as filled circles,
empty circles, and diamonds, respectively.
\label{fig1} }
\end{figure*}

For the five dSphs analyzed in this paper, I obtained the generalized
Plummer profile parameters, $b$ and $\alpha$, by $\chi^2$ fitting of
eq.~(\ref{Sigma}) to the best available star count data: \citet[their
Figure~15]{col05b} for Fornax, \citet[their Figure~9]{smo07} for
Leo~I, \citet[their Figure~6]{col05a} for Sculptor, and \citet{IH95}
for Sextans. For Carina, there were significant discrepancies between
the published star count profiles, so I carried out a joint $\chi^2$
fitting for three different datasets: \citet{IH95}, \citet[their
Table~3]{maj00}, and \citet[their Figure~4b]{mun06}. As the fitting
was done in the linear space, all star count data points (even those
which became formally negative after subtracting the assumed
contribution from the background sources) were used.
Figure~\ref{fig1} shows the star count data and the best fitting
generalized Plummer models for the five dSphs.

\begin{table}
\caption{$\chi^2$ for different surface brightness models\label{tab1}} 
\begin{center}
\begin{tabular}{lcccc}
\tableline
Galaxy     & $\chi^2_{\rm GP}$ & $\chi^2_{\rm P}$ & $\chi^2_{\rm EK}$ & $\chi^2_{\rm S}$ \\
\tableline
Carina   & 2.18              & 2.18             & 2.83              & 2.92\\
Fornax   & 2.35              & 9.81             & 2.86              & 2.55\\
Leo~I    & 3.12              & 47.8             & 11.5              & 6.66\\
Sculptor & 3.50              & 3.50             & 4.86              & 4.98\\
Sextans  & 0.87              & 1.03             & 0.84              & 0.89\\
Draco    & 0.65              & 1.69             & 1.25              & 1.10\\
\tableline
Average  & 2.11              & 11.0             & 4.02              & 3.18\\
\tableline
\end{tabular}
\end{center}
\tablecomments{$\chi^2_{\rm GP}$, $\chi^2_{\rm P}$, $\chi^2_{\rm EK}$, and $\chi^2_{\rm S}$
are normalized $\chi^2$ values for the generalized Plummer, classical
Plummer, empirical King, and S{\'e}rsic profiles, respectively.}
\end{table}

Table~\ref{tab1} presents the best $\chi^2$ values (normalized by the
number of degrees of freedom) for the five dSphs for four different
profiles (generalized Plummer, classical Plummer, empirical King, and
S{\'e}rsic); I also included the fitting results for the Draco data
from \citet[their sample S2]{ode01} -- the same data we used in
\citet{mas06a} for mass modeling of this galaxy. As you can see in
Table~\ref{tab1}, the generalized Plummer profile is the best one
overall: for each galaxy, it produces the same or better fit than
the other profiles, and is significantly better in terms of the average
$\chi^2$ for all the six dwarfs.  I warmly recommend this profile for
all future dSph mass modeling work, as it is simple (surface and
volume densities and gravitational potential are simple analytical
expressions -- as long as $\alpha$ is an integer) and appears to be
the best fitting profile among 2--3 parameter models.

One complication is that all the published star count profiles were
computed inside elliptical annuli (usually with the fixed ellipticity
$e$ and position angle), with the quoted radius measured along
the major axis of the galaxy.   For the purposes of this study,
spherically averaged (in the plane of the sky) surface brightness
profiles are required. For each galaxy, I computed the spherically
averaged profile by solving numerically the following integral
for 100 values of the projected radius $R=0.001\dots 30$:

\begin{equation}
\label{averaging}
\Sigma_{\rm avr}(R) = \frac{2}{\pi}\int\limits_0^{\pi/2} 
\left(1+\frac{R^2}{\cos^2\varphi + (1-e)^2 \sin^2\varphi}\right)^{-\frac{\alpha-1}{2}}{\rm d} \varphi .
\end{equation}

\noindent Here $e$ is the ellipticity of the galaxy. The resultant, averaged profile
is very close (but not identical) to the generalized Plummer profile with
the same exponent $\alpha$ and a smaller core radius $b$. I derived
this spherically averaged value of $b$ by $\chi^2$ fitting the above profile
with the generalized Plummer profile. The correction factor $b/b_{\rm maj}$ 
is a function of both $e$ and $\alpha$, and ranges from 0.816 (Sextans)
to 0.895 (Leo~I). (Here $b_{\rm maj}$ is the core radius obtained from fitting
the generalized Plummer model to the original data, before averaging.)

\begin{table*}
\caption{Input galactic parameters\label{tab2}} 
\begin{center}
\begin{tabular}{lccccccccccc}
\tableline
Galaxy                  & $D$ & $\Sigma_0$        & $e$  & (M/L)$_*$           & $V_{\rm sys}$ & $\alpha$ & $b_{\rm maj}$ & $b$    & $\rho_{*,0}$        & $R_{\rm hl}$\\
                      & kpc & mag arcsec$^{-2}$ &      & M$_\odot$/L$_\odot$ & km s$^{-1}$   &          & arcmin        & arcmin & M$_\odot$ pc$^{-3}$ & arcmin\\
\tableline
Carina                & 101 & 25.5              & 0.33 & 0.845               & 222.9         & 5        & 10.89         & 9.00   & 0.0055              & 3.97\\
Fornax                & 138 & 23.4              & 0.31 & 0.935               & 55.2          & 7        & 27.92         & 23.61  & 0.0147              & 7.67\\
Leo~I                 & 255 & 22.6              & 0.21 & 0.74                & 282.9         & 13       & 10.21         & 9.14   & 0.0490              & 1.92\\
Sculptor              &  79 & 23.7              & 0.32 & 1.23                & 111.4         & 5        & 9.95          & 8.28   & 0.0584              & 3.66\\
Sextans               &  86 & 26.2              & 0.35 & 1.15                & 224.3         & 4        & 22.18         & 18.09  & 0.00195             & 10.45\\
\tableline
\end{tabular}
\end{center}
\tablecomments{Here $D$ is the distance from the Sun, $\Sigma_0$ is the central surface brightness in $V$-band, $e$ is the ellipticity,
(M/L)$_*$ is the stellar mass-to-light ratio in $V$-band, $V_{\rm sys}$ is the systemic heliocentric velocity, $\alpha$ and $b_{\rm maj}$ are
the generalized Plummer parameters, $b$ is the spherically averaged generalized Plummer parameter, $\rho_{*,0}$ is the central stellar density,
and $R_{\rm hl}$ is the half-light radius for a spherically averaged galaxy. $D$, $\Sigma_0$, and $e$ were taken from \citet{mat98} for
all the galaxies except for Leo~I; for Leo~I I used the data from \citet{mat08}. (M/L)$_*$ is from \citet[their Table~6]{mat98b}.
$V_{\rm sys}$ is from \citet{wal09b}. The rest of the parameters were derived in this paper.}
\end{table*}

The results of the generalized Plummer profile fitting and other input
galactic parameters for the five dSphs are presented in
Table~\ref{tab2}. As this study properly takes into account the
self-gravity of stars, I had to assume certain values for the stellar
mass-to-light ratios in $V$-band, (M/L)$_*$. For each galaxy, I used
the average of the Salpeter and composite (M/L)$_*$ estimates of
\citet[their Table~6]{mat98b}. Knowing the distance to the galaxy,
$D$, the $V$-band central surface brightness, $\Sigma_0$, the
generalized Plummer model parameters $\alpha$ and $b$, and (M/L)$_*$
allowed me to compute the central stellar density, $\rho_{*,0}$ (see
Table~\ref{tab2}).  The table also lists the estimated half-light
radii (in arc minutes), $R_{\rm hl}$, for spherically averaged
models.

\subsection{Line-of-sight star velocities}
\label{vlos}

I used the stellar line-of-sight velocities catalogs of \citet{wal09a} for
Carina, Fornax, Sculptor, and Sextans, and the catalog of
\citet{mat08} for Leo~I. These are the largest homogeneous catalogs
available for dwarf spheroidals, with the number of stars per galaxy ranging from $\sim
300$ (Leo~I) to $\sim 2500$ (Fornax). I used the merged version of the
data\footnote{\url{http://www.astro.lsa.umich.edu/$\sim$mmateo/research.html}}.
The median error of the velocity measurements is $\pm
2.1$~km~s$^{-1}$. In all the dwarfs (except for Leo~I) some
contamination from foreground Galactic stars is expected to be present
in the data. \citet{wal09b} quantified this contamination by computing
individual ``membership probabilities'' $P$ for each star. For this
study, I only used stars with $P>0.95$ for Carina, Sculptor, and
Sextans, and stars with $P>0.75$ for Fornax.  (Despite the less
conservative selection criterion for Fornax, the resultant
line-of-sight dispersion profile looks very smooth and
un-contaminated, and the normalized $\chi^2$ for the best fitting model is
$\sim 1$ -- see Figure~\ref{fig2}.) Due to the Leo~I's large heliocentric
velocity and distance from the Sun, the data for this dwarf should have
essentially zero contamination in the range of heliocentric velocities
from 260 to 315~km~s$^{-1}$ \citep[their Fig.~1]{wal07}. Accordingly, I used
these simple velocity cuts to select the Leo~I star candidates.

Carina was the only galaxy in our sample for which the best fitting
model deviated more than two sigmas from the data. This could be due
to unaccounted foreground contamination (which would not be
surprising, given that it is the closest to the Galactic plane dSph
and has a very low surface brightness).  For this galaxy only, I
removed the six stars (after removing the stars with $P<0.95$), for
which the deviation from the best fitting model was more than three
sigmas, and then repeated the full mass modeling for the reduced star
list. (Here one sigma corresponds to the total, model $+$
observational uncertainty, velocity dispersion.) The removed stars
have the following designations in the \citet{wal09a} catalog:
Car-0056, 0200, 0465, 0547, 0680, 1892. This operation boosted the
absolute probability of the best fitting model for Carina from 0.04 to
0.26 (when $N_{\rm bin}=7$; see \S~\ref{method}). It is important to note
that all the six stars are located in the outer half of the radial
bins, so this operation should have had very little impact on the
analysis of the central parts of the galaxy -- which is the main focus of
the current study.

\section{CHOOSING GOOD MODELS}
\label{fitting}

\subsection{$\chi^2$ fitting}
\label{good}

Two popular statistical methods used in mass modeling of dSphs are
$\chi^2$ fitting and maximum likelihood analysis. Both methods have
advantages and disadvantages. The $\chi^2$ method complements
naturally the Jeans analysis, as both operate only with the second
moment (dispersion) of the stellar velocity distribution, whereas for the
maximum likelihood approach one has to assume a certain (usually
Gaussian) form for the line-of-sight velocity probability distribution
function. This factor makes the $\chi^2$ fitting less sensitive
to outliers, and hence a much better approach when one is interested
in absolute probabilities of different models. The maximum likelihood
method, on the other hand, is more appropriate for post-Jeans mass
modeling, where the full model line-of-sight velocity PDF is computed
and compared to the data.

The main disadvantages of the $\chi^2$ technique are that 1) one has
to try a few different values for the number of the radial bins, $N_{\rm bin}$
-- but that only means longer computing time, and 2) in its simplest
implementation, the random locations of individual stars within each
radial bin is an additional source of variance in the analysis. To
address the latter issue, for both the observational data and model I
estimate the following quantity (for each radial bin):

\begin{equation}
\label{sgm_mod}
S=\frac1n\sum_{i=1}^n \sigma_i^2.
\end{equation}

\noindent Here $\sigma_i$ is the line-of-sight velocity dispersion at the
specific location (projected distance from the dwarf's center) of the
$i$-th star, and $n$ is the number of stars in the current radial
bin. The quantity $S$ will converge to the square of the local
line-of-sight velocity dispersion when the radial extent of the
bin shrinks to zero and the number of stars ($n$) grows to
infinity. The important point is that such a convergence is not
required when all one wants to do is to measure the statistical
deviation of the model from the data. All is needed is to estimate the
mean (for the data and the model) and standard deviation (only for the
data) of the quantity $S$ for each radial bin, and then compute the
$\chi^2$ deviation for $S$ from all the radial bins. As the $S$
measurements for both observational data and model are carried out
for the same projected distances from the center of the dwarf
(corresponding to the distances of actual stars), the variance due to
random locations of stars within radial bins is completely removed.

Of course, we do not know from the observations the values of $\sigma_i$ at the position of
each star. Instead, we know the individual line-of-sight velocities
(corrected for the systemic velocity of the galaxy), $V_i$, and the
associated measurement uncertainties, $\sigma_{{\rm mes},i}$. One can
use these known quantities to write down the following unbiased
estimator of $S$ (without any assumptions
on the shape of the velocity PDF):

\begin{equation}
\label{sgm_obs}
\hat S= \frac1n\sum_{i=1}^n \left(V_i^2 - \sigma_{{\rm mes},i}^2 \right).
\end{equation}

\noindent This expression is valid even for the cases when the line-of-sight
velocity dispersion changes dramatically within a single radial bin. The standard
deviation of $\hat S$ can be estimated from the following expression,

\begin{equation}
\label{sgm_std}
\sigma_S\simeq \frac1n \left[2 \sum_{i=1}^n \left(\hat S + \sigma_{{\rm mes},i}^2 \right)^2 \right]^{1/2},
\end{equation}

\noindent which does require the velocity dispersion to be approximately constant within a single bin.
Numerical experiments showed that eq.~(\ref{sgm_std}) noticeably
deviates from the true answer only when the velocity dispersion
changes significantly (by a factor of two or more) within a single
radial bin, which is never the case in my best fitting models.

Eqs.~(\ref{sgm_mod}--\ref{sgm_std}) are written for a single radial
bin. It is now straightforward to write down the expression for the $\chi^2$ 
deviation of the model from the data for all the $N_{\rm bin}$ radial bins:

\begin{equation}
\label{chi2}
\chi^2_k = \sum_{j=1}^{N_{\rm bin}} \left(\frac{\hat S_j-S_j}{\sigma_{S,j}}\right)^2.
\end{equation}

\noindent Here $S_j$, $\hat S_j$, and $\sigma_{S,j}$ are computed for
the $j$-th radial bin using eqs.~(\ref{sgm_mod}), (\ref{sgm_obs}), and
(\ref{sgm_std}), respectively, and $k=N_{\rm bin}-\lambda$ is the
number of the statistical degrees of freedom, where $\lambda$ the
effective dimensionality of the free parameter space (see \S~\ref{test}).

The only drawback of the above procedure is that it can be
substantially more computationally expensive than the standard
$\chi^2$ analysis, as one has to compute the model line-of-sight
velocity dispersion (numerical integration using
eq.~(\ref{sigma_los})) for each model, each star, and each value of
$N_{\rm bin}$.

\subsection{Artificial data tests}
\label{test}

One can already use eq.~(\ref{chi2}) from the previous section to
compute the { \it relative} goodness of fit for different models. To
go one step further and convert the $\chi^2$ values to the {\it
  absolute} probabilities of the models,

\begin{equation}
\label{Prob}
\Pi=P\left(\frac{k}2,\frac{\chi^2_k}2 \right),
\end{equation}

\noindent one has to estimate the effective number of dimensions of
the free parameter space, $\lambda$. (Here $P(x,y)$ is the regularized
Gamma function\footnote{\url{http://en.wikipedia.org/wiki/Chi-square\_distribution}}.)

To estimate $\lambda$ for my six-parameter model described in
\S~\ref{model}, I generated a dataset of 2499 fake stars located at
the same projected distances from the dwarf's center as the real 2499
stars in Fornax with known line-of-sight velocities (see
\S~\ref{vlos}). I used my best fitting Fornax model (for the $N_{\rm
  bin}=7$ case) and actual individual star velocity observational
uncertainties, $\sigma_{\rm mes}$, to generate random values of the
``observed'' line-of-sight velocities for the artificial stars.  (I
assumed that both the model line-of-sight velocities and the
observational velocity uncertainties have Gaussian distributions.)

Next, I split my artificial dataset into 21 equal size sub-sets (each
one consisting of 119 randomly selected fake stars). In a sense, I generated
21 random realizations of the Fornax galaxy. Then I carried out
full mass modeling (as described in \S~\ref{method}, with $N_{\rm
  bin}=7$) and derived the $\chi^2_k$ value (eq.~(\ref{chi2})) for the
best fitting model, separately for each sub-set. Assuming that the
resultant 21 random numbers are drawn from the same $\chi^2_k$
distribution, their mean should be an unbiased estimator for $k$. In
this numerical experiment, I derived $k\simeq 3.73$, corresponding to
the effective number of dimensions in my free parameter space
$\lambda=N_{\rm bin}-k\simeq 3.27$. I used this number in the rest of
the paper.  This number is significantly (almost by a factor of two)
lower than the formal number of the model free parameters (six) reflecting
significant degeneracies (from the Jeans analysis point of view)
present in the model. As one will see in \S~\ref{results}, the most
degenerate is the parameter $\gamma$ (the logarithmic slope of
the DM density).  Post-Jeans mass modeling (dealing with the full
line-of-sight velocity PDF, and not just with the dispersion) is
expected to be less degenerate.

It is clear that $\lambda$ will not be exactly the same for different
data and for different number of radial bins (because the radial
extent of the binned data depends on $N_{\rm bin}$), but as long as
the number of bins is relatively large (so that $k=N_{\rm
bin}-\lambda\gg 1$), the computed absolute probabilities should not
be very sensitive to the errors in $\lambda$. To be more quantitative,
I computed numerically the relative error for the absolute
probabilities $\Pi$ resulting from the $\lambda$ uncertainty,
$(\Delta\lambda / \Pi)\: {\rm d} \Pi/{\rm d} k$, for $k=3.73$ and two
different cases: $\chi^2_k=k$ (corresponding to $\Pi\simeq 0.4$, as in
my best fitting models), and $\chi^2_k=2.5k$ (corresponding to
$\Pi\simeq 0.04$, my lower cutoff for ``good models''). Purely
statistical (Poissonian) errors in estimating $k$ (and hence $\lambda$) from the
limited sample of the size $N_S=21$ is of the order of
$(2k/N_S)^{1/2}\simeq 0.6$.  Assuming a larger uncertainty (to account
for a variance in data) $\Delta\lambda=1$, the resulting relative errors in
$\Pi$ are only $0.05$ and $0.35$, respectively, which I consider to be
acceptable. If $N_{\rm bin}>7$, as is the case with most of my
galaxies, the relative errors in $\Pi$ should be significantly
smaller.

\begin{figure*} 
\epsscale{0.8}
\plotone{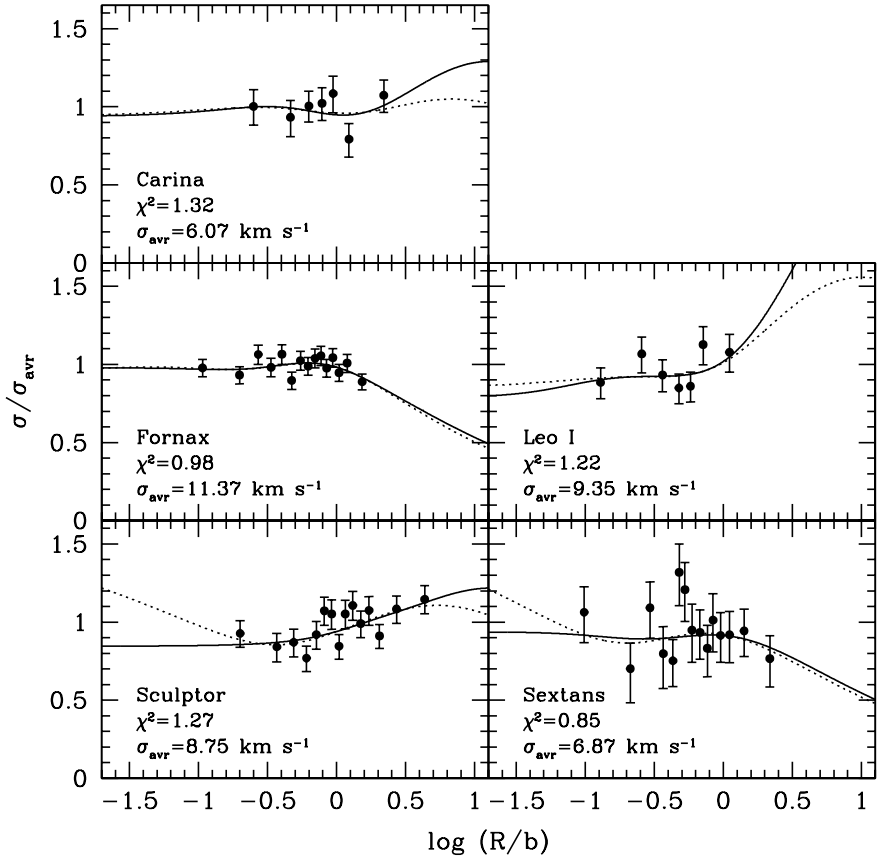} 
\caption {Radial line-of-sight
stellar velocity dispersion profiles.  Solid circles with error bars show the binned
observational data. The solid and dotted lines show the best models with
$\gamma=0$ and 1, respectively.  
\label{fig2} } 
\end{figure*}

\subsection{Numerical algorithm}
\label{method}

In this section I outline my overall mass modeling procedure.

I thoroughly explored (``brute force optimization'') the four out of
the six free model parameters: the central logarithmic DM density slope
$\gamma$, and the three stellar anisotropy parameters, $\eta_0$,
$\eta_1$, and $r_a$ (see \S~\ref{model}). For $\gamma$, I used the
discrete values (0, $0.1$, $0.2$, \dots, $1$), which covered the
whole plausible range -- from the observations-suggested flat core
($\gamma=0$) to the CDM prediction,
$\gamma\sim 1$. In these low stellar density objects the presence
of baryons is not expected to make the DM distribution ``cuspier'' via
the adiabatic contraction of DM. A presence of a sufficiently massive central black
hole could have steepened the DM cusp, but there is no evidence for
super-stellar mass black holes in dSphs \citep{lor09}.

Both $\eta_0$ and $\eta_1$ had the following discrete values: ($-1$, $-0.9$, $-0.8$,
\dots , 1), which covered the full range of anisotropies -- from purely
tangential ($\eta=-1$) to purely radial ($\eta=1$). In addition, for
the central anisotropy $\eta_0$ I only explored the range allowed by
the central anisotropy theorem of \citet{ane06}, $\eta_0\leqslant
(4/\gamma-1)^{-1}$. The anisotropy radius $r_a$ had the following
values (in the units of the stellar core radius $b$): (0.1, 0.147,
0.215, 0.316, 0.464, 0.681, 1.47).  They are equidistant in the
logarithmic space, and span the full range of radial distances
resolved by the data. Overall, I explicitly tested 22,344 different
models with the above discrete values of the four model parameters.

The remaining two model parameters, DM halo scaling radius $r_s$ and
density $\rho_s$, are very well constrained by the data when the rest
of the parameters are fixed, which allowed me to use implicit
optimization technique for these two parameters. This dramatically
reduced the number of models to be tested and hence the total
computational time, as $r_s$ and especially $\rho_s$ span a very large
range of values (four orders of magnitude for $\rho_s$, see
Table~\ref{tab3}). The only prior was for $r_s$ to stay within the
interval 0.05 \dots 5 (in $b$ units; for Sculptor, I used a different
range 0.1 \dots 15, as this galaxy has a very compact core). There was
no prior for $\rho_s$.

I used the one-dimensional Brent's method \citep{NR} separately for $r_s$
and $\rho_s$ optimization (finding global minimum in $\chi^2_k$;
$\rho_s$ optimization is the inner loop, and $r_s$ optimization is the
outer loop). The method showed a good numerical convergence, requiring
only of order of a hundred models with different combinations of
($r_s$, $\rho_s$) to be tested (when the other four explicit
parameters were fixed).  Including both explicitly and implicitly
explored parameters, I had to compute the $\chi^2_k$ deviation from
the data for $\sim 2\times 10^6$ different models -- per galaxy and
per $N_{\rm bin}$ value.

For a given galaxy and a chosen combination of the six model
parameters and the value of $N_{\rm bin}$, I solved the Jeans
eq.~(\ref{Jeans}) numerically, and then solved numerically the
integral in eq.~(\ref{sigma_los}) to derive $\sigma_{\rm los}$ at all
the projected radii $R$ corresponding to the locations of the dwarf's
stars with known line-of-sight velocity. Then I used
eqs.~(\ref{sgm_mod}--\ref{chi2}) to compute the $\chi^2$ deviation of
the model from the data, and eq.~(\ref{Prob}) to assign absolute
probabilities to the models.

Absolute probabilities of the best fitting models, $\Pi_{\rm max}$,
range from 0.23 for Sculptor to 0.60 for Sextans (Table~\ref{tab3}),
corresponding to $\sim 0.5-1.2$ sigma deviations. I used
one global lower cut-off value of $\Pi_{\rm low}=0.0455$ (a
two sigma deviation) to select good models for all the galaxies.

I used the following approach to determine the optimal value of the
number of radial bins, $N_{\rm bin}$, for each galaxy. In the limit
of large $N_{\rm bin}$ (small number of stars in one bin, $n$),
$\sigma_S\propto N_{\rm bin}^{1/2}$ (see eq.(\ref{sgm_std})) and the
$\chi^2$ analysis becomes dominated by the Poissonian noise and hence
``fuzzier'' (less discriminatory).  In the opposite limit, $N_{\rm
bin}\rightarrow 0$, the signal to noise ratio improves, but the
$\chi^2$ analysis becomes again less accurate because with fewer
radial bins we start loosing radial resolution and may overlook some
sharp $\sigma_{\rm los}$ features near the dwarf's center or in the
galactic outskirts. In addition, when $N_{\rm bin}$ approaches
$\lambda$ (the effective dimensionality of the free parameter space),
the errors in measuring $\lambda$ become more important, which makes
the absolute model probabilities less reliable (see
\S~\ref{test}). Hence it appears there should be a certain value of
$N_{\rm bin}$ when the $\chi^2$ analysis is the most accurate (or
discriminatory). I found this optimal value separately for each galaxy
by running the full set of models for $N_{\rm bin}=7$ (minimum
acceptable number of bins for this model, see \S~\ref{test}), 15, and
30, and then choosing the value which produced the smallest fraction
of good models ($f_{2\sigma}$ in Table~\ref{tab3}). For Fornax,
Sculptor, and Sextans the optimal value of $N_{\rm bin}$ was found to
be 15; for Carina and Leo~I it is equal to 7 (Table~\ref{tab3}).

\section{RESULTS}
\label{results}

I applied the mass modeling procedure described in the previous
sections to all the five dwarfs. The full results are available
online. As good models found in the current analysis occupy only a
small fraction of the total six-dimensional free parameter space, the
data presented here can be used for post-Jeans mass modeling projects
(utilizing the full PDF of the observed stellar line-of-sight
velocities; the subject of Paper~II) to dramatically reduce the
required computational time.

Figure~\ref{fig2} shows the radial $\sigma_{\rm los}$ profiles for the two
best-fitting models (with $\gamma=0$, solid lines, and $\gamma=1$,
dotted lines) for each galaxy. As one can see, the overall quality of
the model fits is excellent, with the $\chi^2$ values (normalized by
the number of degrees of freedom $k=N_{\rm bin}-\lambda$) being close
to a unity.  It is remarkable that these two dramatically different
values of $\gamma$ (0 and $1$) make very little difference in the
quality of the model fit.  This is most obvious for the galaxy with
the highest quality data -- Fornax -- where the two $\sigma_{\rm los}$
profiles are essentially identical way beyond the radial range covered
by the data. For the galaxies with the best $\gamma=0$ and $\gamma=1$
profiles apparently diverging near the center (Sculptor and Sextans),
one could na\"ively think that with a larger $N_{\rm bin}$ (and hence
with the binned data covering a wider range of radial distances) the two
models would become more statistically distinct. But this is not the
case: for all the values of $N_{\rm bin}$ I tried (7, 15, and 30), and
for all the five galaxies, the difference between the best fitting
models with different values of $\gamma$ is statistically
insignificant, to a similar degree.

\begin{table*}
\caption{Global constraints\label{tab3}} 
\begin{center}
\begin{tabular}{lccccccccccccccccccc}
\tableline
Name     &  $N_*$ & $N_{\rm bin}$ &$f_{2\sigma}$&$\Pi_{\max}$& $\eta_0$        &$r_s$           & $\rho_s$                   & $\rho_{0,{\rm min}}$ &$(M/L)_{0,{\rm min}}$ \\
         &        &               &             &            &                 &                &                            & $M_\odot$~pc$^{-3}$  & $M_\odot / L_\odot$  \\
\tableline	           
Carina   &  740   &  7            &  0.680      & 0.262      & $-1\dots 0.3$   & $0.05\dots 5$  & $1.7\dots 2.5\times 10^4$  & 0.048                & 8.2                  \\
Fornax   & 2499   & 15            &  0.253      & 0.468      & $-0.8\dots 0.3$ & $0.05\dots 5$  & $0.25\dots 2.1\times 10^3$ & 0.028                & 2.7                  \\
Leo I    &  328   &  7            &  0.210      & 0.302      & $-0.8\dots 0.3$ & $0.05\dots 5$  & $0.13\dots 600$            & 0.080                & 1.95                 \\
Sculptor & 1349   & 15            &  0.230      & 0.228      & $-0.9\dots 0.3$ & $0.5\dots 15$  & $0.16\dots 63$             & 0.087                & 3.1                  \\
Sextans  &  397   & 15            &  0.906      & 0.598      & $-1\dots 0.3$   & $0.05\dots 5$  & $0.90\dots 2.9\times 10^4$ & 0.007                & 5.3                  \\
\tableline
\end{tabular}
\end{center}
\tablecomments{Here $N_*$ is the number of stars with a known line-of-sight velocity used for the analysis; 
$N_{\rm bin}$ is the number of radial bins; $f_{2\sigma}$ is the fraction of the ``good models'' (with the absolute
probability $\Pi>0.0455$); $\Pi_{\max}$ is the highest absolute probability for a model. I also list
the two-sigma ranges spanned by good models for the global model parameters $\eta_0$ (central anisotropy),
$r_s$ and $\rho_s$ (DM scaling radius and density). Finally, the two-sigma constraints on the 
minimum central DM density ($\rho_{0,{\rm min}}$) and the corresponding mass to light ratio ($(M/L)_{0,{\rm min}}$)
are given.
}
\end{table*}

\begin{table*}
\caption{Radius-dependent constraints\label{tab5}} 
\begin{center}
\begin{tabular}{lccccccccccccccccccc}
\tableline
Name     & $r_1$ & $M_1$            & $(M/L)_1$          & $r_2$ & $\rho_{\rm DM}$     &$(M/L)_2$            & $r_3$ & $\eta$            & $M_{300}$      & $M_{600}$ \\
         & pc    & $10^7 M_\odot$   &$M_\odot / L_\odot$ & pc    & $M_\odot$~pc$^{-3}$ & $M_\odot / L_\odot$ & pc    &                   & $10^7 M_\odot$ & $10^7 M_\odot$ \\
\tableline                                                                                                           
Carina   & 409   & 0.74\dots 1.14   & 26\dots 39         & 187   & 0.032\dots 0.074    & 14\dots 32          & \dots & $-1$\dots 0.3     & 0.35\dots 0.78 & 0.85\dots 2.3\\
Fornax   & 924   & 6.1\dots 8.0     & 5.5\dots 6.9       & 453   & 0.021\dots 0.040    & 3.7\dots 6.2        & 232   & $-0.88$\dots 0.05 & 0.27\dots 3.4  & 1.90\dots 5.1\\
Leo I    & 389   & 1.55\dots 1.94   & 3.4\dots 4.1       & 204   & 0.067\dots 0.115    & 2.5\dots 3.8        & 72    & $-0.67$\dots 0.26 & 0.76\dots 1.43 & 1.94\dots 5.7\\
Sculptor & 435   & 2.17\dots 3.01   & 22\dots 30         & 256   & 0.067\dots 0.107    & 20\dots 31          & 83    & $-0.46$\dots 0.19 & 0.79\dots 1.63 & 2.9\dots 5.8\\
Sextans  & 1034  & 1.4\dots 3.2     & 19\dots 42         & 451   & 0.004\dots 0.013    & 11\dots 32          & \dots & $-1$\dots 0.3     & 0.06\dots 1.1  & 0.37\dots 1.7\\
\tableline
\end{tabular}
\end{center}
\tablecomments{
  Here $r_1$ is the radius where the enclosed DM mass has the tightest constraints. 
  At this radius, I list the two-sigma constraints on the enclosed mass ($M_1$) and the averaged light to mass ratio ($(M/L)_1$).
  Similarly, $r_2$ is the radius where the local DM density has the tightest constraints, with the corresponding
  two-sigma constraints on the local DM density ($\rho_{\rm DM}$) and the local mass to light ratio ($(M/L)_2$).
  Finally, $r_3$ is the radius where the local stellar velocity anisotropy, $\eta$, has the tightest constraints.
  In addition, I show the two-sigma constraints for the enclosed mass at the radii 300~pc ($M_{300}$) and 600~pc ($M_{600}$).
}
\end{table*}

Tables~\ref{tab3} and \ref{tab5} summarize the main results of this study. In
particular, I list there the ranges spanned by good ($<2\sigma$)
models separately for the free model parameters $\eta_0$, $r_s$, and
$\rho_s$. As one can see, these parameters are very poorly constrained
by the Jeans analysis. The full allowed range for $\eta_0$ is $-1\dots 0.3$
(due to the \citealt{ane06} central anisotropy theorem), and only three
galaxies (Fornax, Leo~I, and Sculptor) have a (very weak) constraint
on this parameter, in the sense that the pure central tangential
anisotropy is excluded. Similarly, only one galaxy (Sculptor) has any
constraints on the DM halo scaling radius: $r_s\gtrsim 0.5$ (in $b$
units). For the rest of the dwarfs, good models span the full
explored range of $r_s$, from 0.05 to 5. Parameter $\rho_s$ is the
only one in the current analysis which is allowed to take any values
(no priors), so it does appear to be constrained in a meaningful
way. One has to remember though that there is a strong degeneracy
between $\rho_s$ and $r_s$. E.g., if $r_s$ were allowed to go to even
smaller values ($<0.05$), the range for $\rho_s$ would expand to even
larger values. I do not list global constraints for the three
remaining model parameters, $\gamma$, $\eta_1$, and $r_a$, as good
models span the full ranges for these parameters: $0\dots 1$, $-1
\dots 1$, and $0.1 \dots 1.47$, respectively. The conclusion here is
that in mass modeling of dwarf spheroidal galaxies the Jeans analysis
is not capable of placing meaningful  constraints on global  model
parameters (for models which are flexible enough, as in the current
study).

The analysis does constrain the minimum central DM density,
$\rho_{\rm 0,\rm min}$, and the corresponding mass to light ratio,
$(M/L)_{0,\rm min}$ (see Table~\ref{tab3}).  Sculptor and Leo~I have the
strongest constraints on $\rho_{\rm 0,\rm min}$ ($\gtrsim
0.09$~M$_\odot$~pc$^{-3}$ and $\gtrsim 0.08$~M$_\odot$~pc$^{-3}$,
respectively, at the two-sigma level). Carina, on the other hand, has
the strongest $(M/L)_{0,\rm min}$ constraint:
$\gtrsim 8$~M$_\odot/$L$_\odot$.

\begin{figure*}
\epsscale{0.8}
\plotone{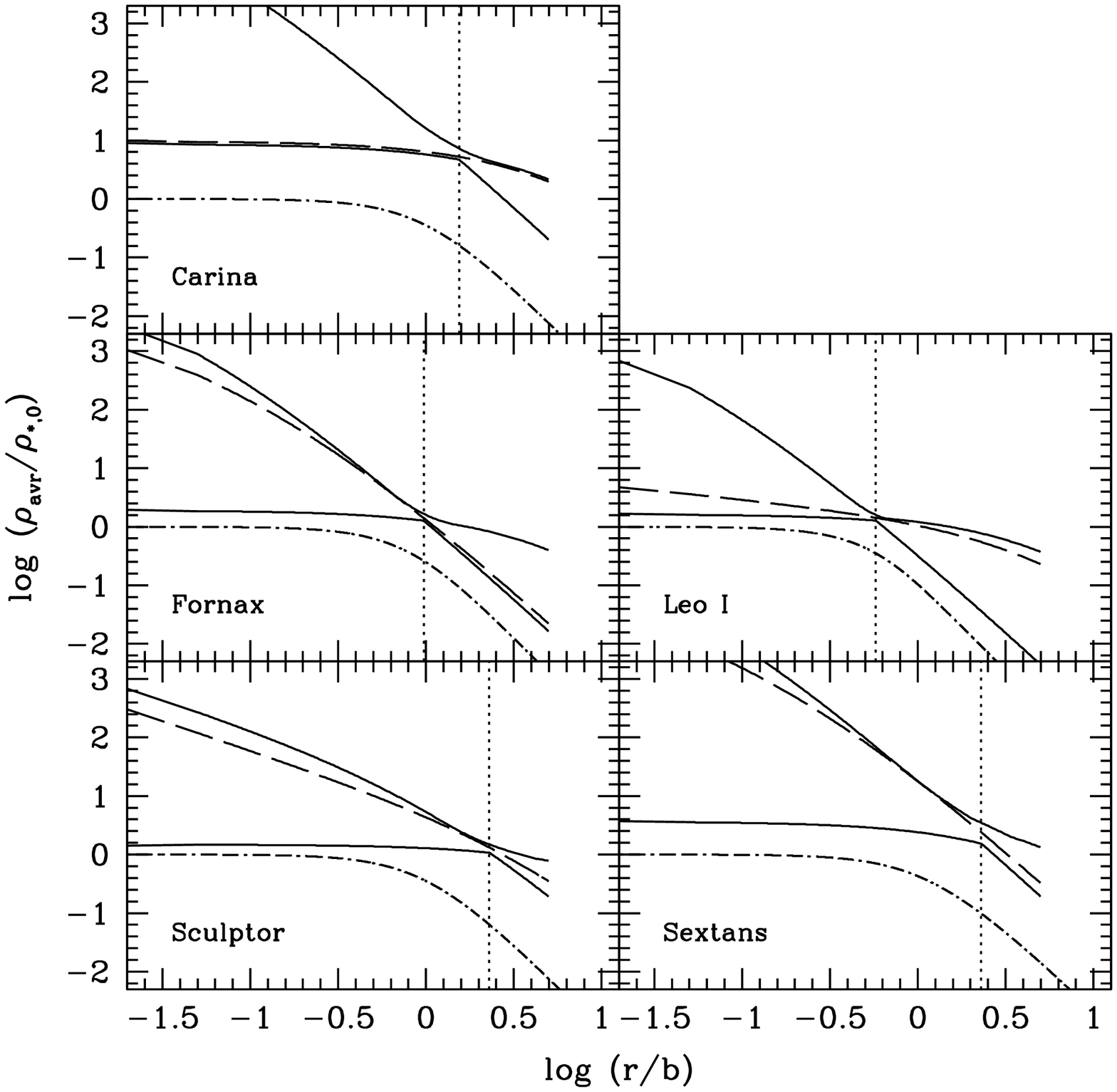}
\caption {Averaged DM density, $\rho_{\rm avr}$, as a function of radius, $r$, 
in stellar units ($\rho_{*,0}$ and $b$, respectively).
Two solid lines outline the range of the models consistent with the
observational data at better than $2\sigma$ level. Dashed lines correspond to
formally best models. Vertical dotted lines mark the radius ($r_1$ in Table~\ref{tab3}) where
the $\rho_{\rm avr}$ uncertainty is the smallest. Dash-dotted lines
show the averaged density profiles for stars.
\label{fig3} }
\end{figure*}

More useful constraints can be derived for radius-dependent (computed
at different distances from the dwarf's center) quantities.
Figure~\ref{fig3} plots the range of the averaged DM density
$\rho_{\rm avr}$ spanned by good models as a function of the enclosed
radius $r$, separately for each galaxy. The averaged stellar density
profiles (dash-dotted lines) are shown for comparison. There has been
a debate in the literature \citep{mat93,str07,str08,wal09c,wol09}
regarding the value of the radius at which the Jeans analysis provides
the tightest constraints on the enclosed mass (or equivalently the
averaged density). As one can see in Figure~\ref{fig3}, such radius
indeed exists (vertical dotted lines) and is very well defined. In
Table~\ref{tab5} I list these radii ($r_1$), the range of the enclosed
mass spanned by good models at those radii ($M_1$), and also the
corresponding range of the mass-to-light ratio ($(M/L)_1$), for all
the dwarfs. For comparison, in Table~\ref{tab5} I also list the good
model ranges of the enclosed mass at 300 and 600~pc -- $M_{300}$ and
$M_{600}$, respectively. As one can see, $r_1$ ranges widely, from
390~pc for Leo~I to more than 1000~pc for Sextans, and at neither of
the previously suggested radii (300 and 600~pc) the enclosed mass is
well constrained by the Jeans analysis for all the dwarfs. To be more
quantitative, the mean-squared two-sigma error for all the five dwarfs
is $\pm 0.40$~dex, $\pm 0.24$~dex, and $\pm 0.10$~dex ($\pm 25\%$) for
the enclosed mass measured within 300~pc, 600~pc, and $r_1$
(unique for each galaxy), respectively. The accurately determined
values of the enclosed mass presented here can be very useful for
identifying the dSph counterparts in cosmological simulations, and is probably
the most important result of the current study.

The conjecture of \citet{wal09c} that the enclosed mass
is well constrained at the half-light radius, $R_{\rm hl}$, is
not corroborated by the current study. In fact, enclosed masses
measured at this radius result in the worst mean-squared two-sigma error, 
$\pm 0.58$~dex.

The conclusion of \citet{wol09} (whose analysis is hard to interpret,
as DM and stars are coupled in their model) that the enclosed mass has
the best constraint at the radius where $\gamma_*\equiv -{\rm d} \log
\rho_*/{\rm d} \log r=3$ is also not borne out in the current study: for
the five dwarfs I obtained $\gamma_*=3.54 \pm 0.38$ at my best radii
($r_1$), with the full range from 3.22 to 4.20.  My sample of galaxies is too
small to try to find a good estimator for $r_1$, but I believe it
should be a function of both the stellar density profile
($\gamma_*(r)$) and the quality and quantity of the stellar
line-of-sight velocity measurements. This issue needs to be explored
further once a larger sample of dwarfs is subjected to a similar
analysis.

To estimate how sensitive the above analysis is to particular
values of $N_{\rm bin}$, I repeated it for all the three values of
$N_{\rm bin}$: 7, 15, and 30. For the less optimal values of $N_{\rm
  bin}$, the radii where the enclosed mass is measured to the highest
possible accuracy ($r_1$) does not deviate from the value obtained
with the optimal $N_{\rm bin}$ by more than 7\% (for Sextans; for
Fornax it is essentially zero). The averaged DM density measured at
$r_1$ is also very close for all the values of $N_{\rm bin}$ (less
than one sigma deviation), with one notable exception: for Leo~I,
moving from $N_{\rm bin}=7$ (the optimal value) to $N_{\rm bin}=30$
results in $\rho_{\rm avr}(r_1)$ becoming smaller by 0.18~dex, or
$3.8\sigma$. (This may be related to the fact that Leo~I has the smallest
number of stars with a known line-of-sight velocity: $N_*=328$.) Overall, this analysis seems to be largely
insensitive to how well the optimal value of $N_{\rm bin}$ is chosen.

Another proposition discussed in the literature is that there is a
radius (300~pc) at which the enclosed mass is the same for all dwarf
spheroidals -- around $10^7$~M$_{\odot}$ \citep{mat93,str08}. This is barely (at
$2\sigma$ level) consistent with the data in the current analysis, as
there is a tension between Carina on one side, and Leo~I and Sculptor
-- on the other side (see Table~\ref{tab5}). The galaxies may have a
similar mass at this radius, but it is unlikely to be identical. If
one considers even smaller radii (say, 200~pc), then formally there will
be a common enclosed mass for all the five dwarfs consistent with the data at
better than two sigma level, but only because the uncertainties in
measuring the enclosed mass increase dramatically at smaller radii
(see Figure~\ref{fig3}).

\begin{figure*}
\epsscale{0.8}
\plotone{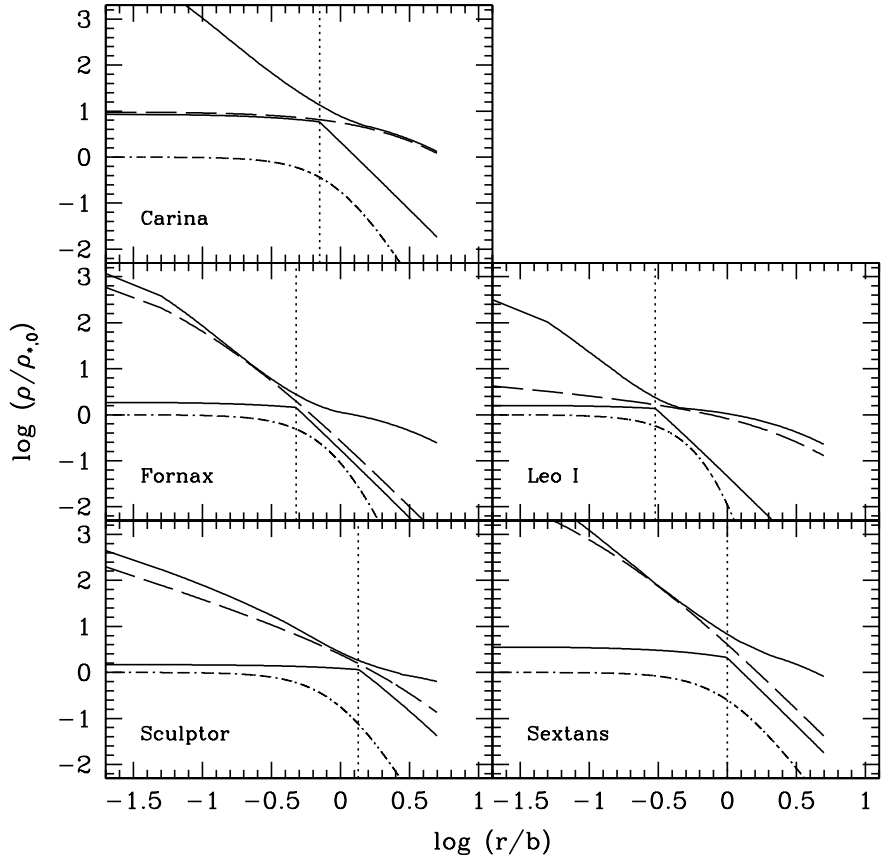}
\caption {Local DM density, $\rho$, as a function of radius, $r$, 
in stellar units ($\rho_{*,0}$ and $b$, respectively).
Two solid lines outline the range of the models consistent with the
observational data at better than $2\sigma$ level. Dashed lines correspond to
formally best models. Vertical dotted lines mark the radius  ($r_2$ in Table~\ref{tab3}) where
the $\rho$ uncertainty is the smallest. Dash-dotted lines
show the density profiles for stars.
\label{fig4} }
\end{figure*}

Another interesting radius-dependent quantity, local DM density,
is more model dependent, and is not as well constrained by the Jeans
analysis as the enclosed mass. Figure~\ref{fig4} shows the ranges for
$\rho_{\rm DM}(r)$ spanned by good models. I also plot there the
stellar density profiles, $\rho_{*}(r)$, as dash-dotted lines. As with
the averaged density case, each galaxy has a well defined radius, $r_2$,
(vertical dotted lines) where the local DM density is best constrained
by the data. Table~\ref{tab5} lists the values of $r_2$ and the
corresponding ranges of $\rho_{\rm DM}(r_2)$ and the local mass-to-light
ratio $(M/L)_2$ for each galaxy. $r_2$ tends to be a factor of two
smaller than $r_1$. The uncertainty in $\rho_{\rm DM}(r_2)$ ranges
from 0.20~dex for Sculptor to 0.5~dex for Sextans, which is
substantially worse than the constraints on the enclosed mass. Still,
knowing the local DM density at a certain radius (which happens to be
quite small) to a reasonably good accuracy can be very useful -- for
example, for studies aimed at direct search of DM in dwarf spheroidals
via its annihilation signal. Both  Figures~\ref{fig3} and \ref{fig4} clearly
demonstrate that these dSphs are DM dominated at all radii.

\begin{figure*}
\epsscale{0.8}
\plotone{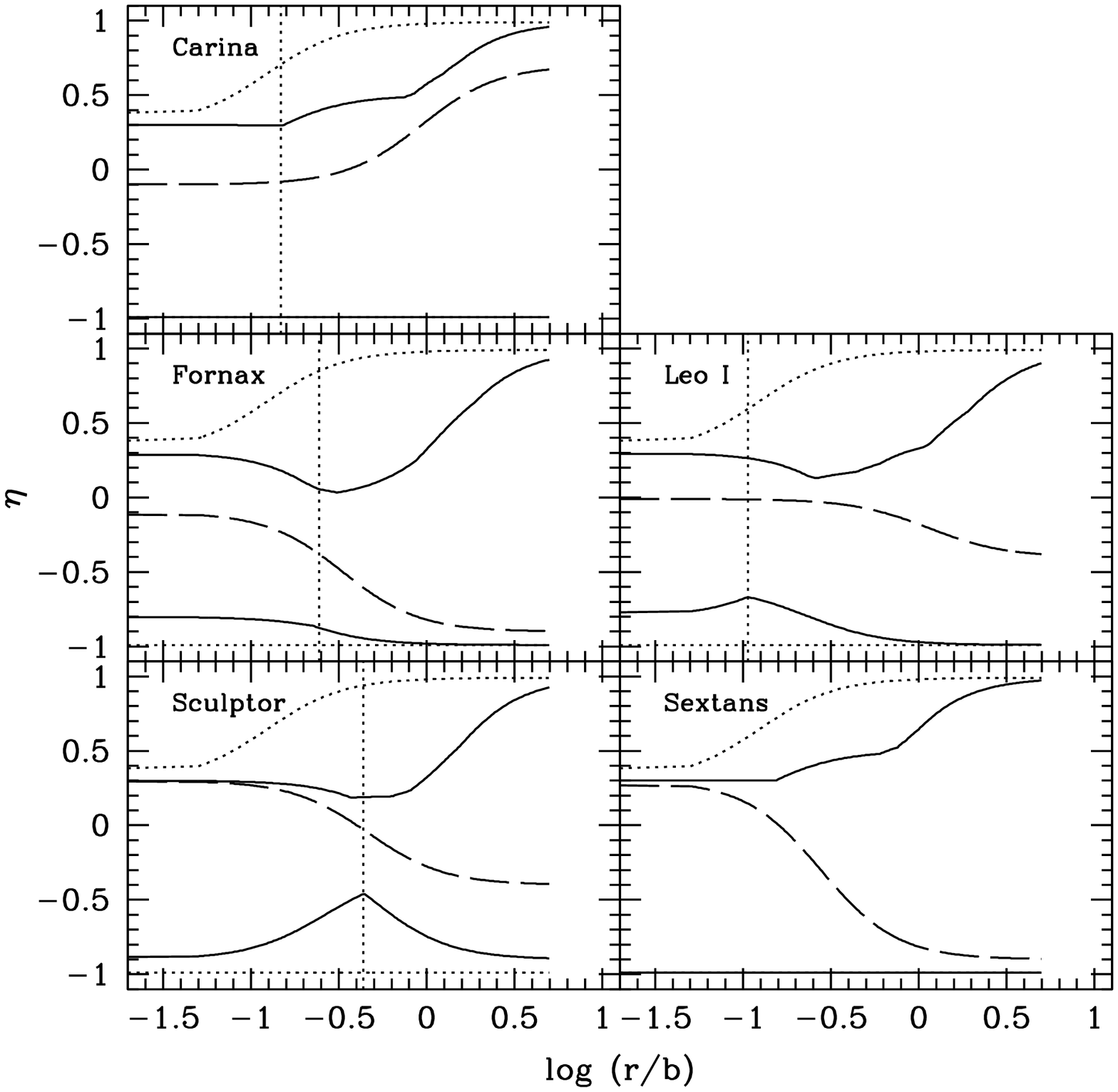}
\caption {Stellar anisotropy, $\eta$, as a function of radius, $r$ (in units of $b$).
Two solid lines outline the range of the models consistent with the
observational data at better than $2\sigma$ level. Two dotted lines show the total
allowed range of models. Dashed lines correspond to formally best
models. Vertical dotted lines mark the radius ($r_3$ in
Table~\ref{tab3}) where the $\eta$ uncertainty is the smallest.
\label{fig5} }
\end{figure*}

The last radius-dependent quantity which has useful (albeit very weak)
constraints is the local stellar velocity anisotropy, $\eta(r)$.
Figure~\ref{fig5} demonstrates that better than two sigma models span
a wide range of $\eta$ at each radius $r$, which is only slightly
narrower than the full allowed range (the space between the two dotted
lines).  The best radius, $r_3$, is not as well defined as with
$\rho_{\rm avr}(r)$ and $\rho_{\rm DM}(r)$.  I list the values of
$r_3$ and the corresponding good model ranges for $\eta(r_3)$ in
Table~\ref{tab5}.  Sculptor has the tightest constraints on
$\eta(r_3)$; Carina and Sextans are essentially unconstrained.

Given how poor the constraints on $\eta(r)$ are, very little can be
learned from the analysis of Figure~\ref{fig5}. One interesting
conclusion is that at any given radius the data are consistent with
stellar velocities being isotropic ($\eta(r)=0$). One cannot conclude
from this radius-dependent analysis that the {\it global} stellar isotropy is
also consistent with the data.  To address the latter issue, one has
to analyze the full results of the current study (available online).
Specifically, one has to test if all the globally isotropic models
(with $\eta_0=\eta_1=0$) are good ones (deviate less than $2\sigma$
from the data). I carried out such an analysis, with the conclusion
being that indeed the data is consistent (at better than $2\sigma$
level) with the global stellar velocity isotropy hypothesis for all
the dwarfs.

It has already been demonstrated \citep{str08,wal09c} that in dSphs
the central logarithmic DM density slope, $\gamma$, is not constrained
by the Jeans analysis (when no stellar proper motion data are available).
My analysis generalizes this result for all radii $r$: the local
logarithmic DM density slope, $\gamma(r)=-{\rm d} \log \rho_{\rm
  DM}/{\rm d} \log r$, is found to be unconstrained by the Jeans analysis.

\section{CONCLUSIONS}
\label{conclusions}

I carried out the classical Jeans mass modeling of the five dwarf
spheroidal galaxies with the highest quality observational data --
Carina, Fornax, Leo~I, Sculptor, and Sextans. My primary goal was to
push the analysis to its limits, by developing a flexible enough
galactic model (with variable stellar velocity anisotropy), refining
the traditional $\chi^2$ fitting algorithm, and investing a
significant amount ($3\times 10^5$~cpu~hours) of supercomputing time
to carry out an exhaustive model parameter optimization. The main
results of this study are as follows.

\begin{itemize}

\item My galactic model with the six free parameters ($r_s$, $\rho_s$,
  $\gamma$, $\eta_0$, $\eta_1$, and $r_a$) gives a good description of
  the observational data: the normalized $\chi^2$ for the differences
  between the model and observed velocity dispersion profiles are
  close to a unity, which is consistent with the deviations being
  purely due to the observational errors.

\item I show that the Jeans mass modeling approach (even with high
  quality data and modeling) cannot place meaningful constraints on
  most galactic parameters, including the central DM density
  logarithmic slope ($\gamma$), DM halo scaling radius ($r_s$) and
  density ($\rho_s$), and central stellar velocity anisotropy
  ($\eta_0$). Moreover, I show that the local DM density logarithmic
  slope, $\gamma(r)=-{\rm d} \log \rho_{\rm DM}/{\rm d} \log r$, is
  unconstrained at all the radii probed by the data.  As a consequence,
  one must resort to much more computationally expensive post-Jeans
  mass modeling techniques (which deal with the full probability
  distribution function for stellar line-of-sight velocities, and not
  just the dispersion) to be able to solve the ``cusp -- core'' issue
  for dSphs, or to identify theoretical mechanisms responsible for
  flattening DM cusps in dwarf galaxies (by constraining the stellar
  velocity anisotropy profiles).

\item My most important finding is that there is a certain radius
  ($r_1$; different for each galaxy) where Jeans mass modeling
  provides the tightest constraints on the enclosed mass in dSphs. For
  Carina, Fornax, Leo~I, Sculptor, and Sextans my two-sigma
  constraints for the enclosed DM mass are $0.74\dots 1.14\times
  10^7$~M$_\odot$ (at 409~pc), $6.1\dots 8.0\times 10^7$~M$_\odot$ (at
  924~pc), $1.55\dots 1.94\times 10^7$~M$_\odot$ (at 389~pc),
  $2.17\dots 3.01\times 10^7$~M$_\odot$ (at 435~pc), and $1.4\dots
  3.2\times 10^7$~M$_\odot$ (at 1034~pc), respectively. These
  constraints are much tighter than the constraints at the previously
  suggested ``good'' radii -- 300~pc, 600~pc, and the half-light
  radius. The tight constraints can be very valuable for placing the
  dSphs in the proper cosmological context.

\item I also derive useful constraints on the local DM density.
  Similarly to the enclosed mass, each galaxy has a certain radius
  ($r_2$) where my Jeans analysis places the tightest constraints on
  this quantity. For example, I show that in Sculptor $\rho_{\rm
    DM}=0.067 \dots 0.107$~$M_\odot$~pc$^{-3}$ (two-sigma interval) at
  256~pc from its center. Also, the analysis produced useful minimum
  central DM density ($\rho_{0,{\rm min}}$) constraints (the largest
  one is for Sculptor: $\rho_0\gtrsim 0.09$~$M_\odot$~pc$^{-3}$).
  These constraints can be used in projects aimed at
  detecting DM in dSphs via its annihilation signal.

\item I show that stellar anisotropy profiles are very poorly
  constrained in Jeans mass modeling. The only useful results here are
  that the pure central tangential anisotropy is excluded at better than
  two-sigma level for three out of the five dSphs, and that the data
  are consistent with the global stellar velocity isotropy for all the five
  dwarfs.

\item A significant advantage of the exhaustive search through the
  multi-dimensional model parameter space employed in this study, with
  all the intermediate results stored and available online, is that
  the results can be used for many other projects. Most importantly,
  this analysis can be used as the first preliminary step (which
  eliminates a vast majority of models which are incompatible with the
  data) in post-Jeans mass modeling projects (which model the full PDF
  for the stellar line-of-sight velocities) -- the subject of my
  Paper~II in this series. The latter approach should be able to
  overcome most of the degeneracies present in Jeans mass modeling (as
  exposed in the current work), hopefully settling once and for all
  whether dSphs have flat DM cores.

\end{itemize} 

\acknowledgements

The simulations reported in this paper were carried out on facilities
of the Shared Hierarchical Academic Research Computing Network
(SHARCNET:www.sharcnet.ca). This research was supported in part by
SHARCNET.


\begin{thebibliography}

 \bibitem[Aaronson(1983)]{aar83} Aaronson, M.\ 1983, \apjl, 266, L11 
 \bibitem[An \& Evans(2006)]{ane06} An, J.~H., \& Evans, N.~W.\ 2006, \apj, 642, 752 
 \bibitem[Baes \& van Hese(2007)]{bae07} Baes, M., \& van Hese, E.\ 2007, \aap, 471, 419 
 \bibitem[Binney \& Tremaine(1987)]{BT} Binney, J., \& Tremaine, S. 1987, Galactic Dynamics (Princeton: Princeton Univ. Press)
 \bibitem[Burkert(1995)]{bur95} Burkert, A.\ 1995, \apjl, 447, L25
 \bibitem[Coleman et al.(2005a)]{col05a} Coleman, M.~G., Da Costa, G.~S., \& Bland-Hawthorn, J.\ 2005, \aj, 130, 1065 
 \bibitem[Coleman et al.(2005b)]{col05b} Coleman, M.~G., Da Costa, G.~S., Bland-Hawthorn, J., \& Freeman, K.~C.\ 2005, \aj, 129, 1443 
 \bibitem[de Blok \& Bosma(2002)]{deb02} de Blok, W.~J.~G., \& Bosma, A.\ 2002, \aap, 385, 816 
 \bibitem[El-Zant et al.(2001)El-Zant, Shlosman, \& Hoffman]{elz01} El-Zant, A., Shlosman, I., \& Hoffman, Y.\ 2001, \apj, 560, 636
 \bibitem[Gentile et al.(2005)]{gen05} Gentile, G., Burkert, A., Salucci, P., Klein, U., \& Walter, F.\ 2005, \apjl, 634, L145 
 \bibitem[Goerdt et al.(2006)]{goe06} Goerdt, T., Moore, B., Read, J.~I., Stadel, J., \& Zemp, M.\ 2006, \mnras, 368, 1073 
 \bibitem[Irwin \& Hatzidimitriou(1995)]{IH95} Irwin, M., \& Hatzidimitriou, D.\ 1995, \mnras, 277, 1354 
 \bibitem[King(1962)]{kin62} King, I.\ 1962, \aj, 67, 471 
 \bibitem[Lora et al.(2009)]{lor09} Lora, V., S{\'a}nchez-Salcedo, F.~J., Raga, A.~C., \& Esquivel, A.\ 2009, \apjl, 699, L113
 \bibitem[Majewski et al.(2000)]{maj00} Majewski, S.~R., Ostheimer, J.~C., Patterson, R.~J., Kunkel, W.~E., Johnston, K.~V., \& Geisler, D.\ 2000, \aj, 119, 760 
 \bibitem[Marchesini et al.(2002)]{mar02} Marchesini, D., D'Onghia, E., Chincarini, G., Firmani, C., Conconi, P., Molinari, E., \& Zacchei, A.\ 2002, \apj, 575, 801 
 \bibitem[Martin et al.(2007)]{mar07} Martin, N.~F., Ibata, R.~A., Chapman, S.~C., Irwin, M., \& Lewis, G.~F.\ 2007, \mnras, 380, 281 
 \bibitem[Mashchenko \& Sills(2004)]{mas04} Mashchenko, S., \& Sills, A.\ 2004, \apjl, 605, L121 
 \bibitem[Mashchenko \& Sills(2005)]{mas05} Mashchenko, S., \& Sills, A.\ 2005, \apj, 619, 243 
 \bibitem[Mashchenko et al.(2006a)Mashchenko, Sills, \& Couchman]{mas06a} Mashchenko, S., Sills, A., \& Couchman, H.~M.\ 2006, \apj, 640, 252
 \bibitem[Mashchenko et al.(2006b)Mashchenko, Couchman, \& Wadsley]{mas06b} Mashchenko, S., Couchman, H.~M.~P., \& Wadsley, J.\ 2006, \nat, 442, 539 
 \bibitem[Mashchenko et al.(2008)Mashchenko, Wadsley, \& Couchman]{mas08} Mashchenko, S., Wadsley, J., \& Couchman, H.~M.~P.\ 2008, Science, 319, 174 
 \bibitem[Mateo(1998)]{mat98} Mateo, M.~L.\ 1998, \araa, 36, 435 
 \bibitem[Mateo et al.(1993)]{mat93} Mateo, M., Olszewski, E.~W., Pryor, C., Welch, D.~L., \& Fischer, P.\ 1993, \aj, 105, 510 
 \bibitem[Mateo et al.(1998)]{mat98b} Mateo, M., Olszewski, E.~W., Vogt, S.~S., \& Keane, M.~J.\ 1998, \aj, 116, 2315 
 \bibitem[Mateo et al.(2008)]{mat08} Mateo, M., Olszewski, E.~W., \& Walker, M.~G.\ 2008, \apj, 675, 201 
 \bibitem[Merritt(1985)]{mer85} Merritt, D.\ 1985, \aj, 90, 1027
 \bibitem[Moore et al.(1999)]{moo99} Moore, B., Ghigna, S., Governato, F., Lake, G., Quinn, T., Stadel, J., \& Tozzi, P.\ 1999, \apjl, 524, L19 
 \bibitem[Mu{\~n}oz et al.(2006)]{mun06} Mu{\~n}oz, R.~R., et al.\ 2006, \apj, 649, 201 
 \bibitem[Navarro et al.(1997)Navarro, Frenk, \& White]{NFW} Navarro, J.~F., Frenk, C.~S., \& White, S.~D.~M.\ 1997, \apj, 490, 493 
 \bibitem[Odenkirchen et al.(2001)]{ode01} Odenkirchen, M., et al.\ 2001, \aj, 122, 2538 
 \bibitem[Osipkov(1979)]{osi79} Osipkov, L. P. 1979, Pis'ma Astron. Zh., 5, 77
 \bibitem[Press et al.(1992)]{NR} Press, W.~H., Teukolsky, S.~A., Vetterling, W.~T., \& Flannery, B.~P.\ 1992, Cambridge: University Press, |c1992, 2nd ed.,  
 \bibitem[S{\'e}gall et al.(2007)]{seg07} S{\'e}gall, M., Ibata, R.~A., Irwin, M.~J., Martin, N.~F., \& Chapman, S.\ 2007, \mnras, 375, 831 
 \bibitem[S{\'e}rsic(1963)]{ser63} S{\'e}rsic, J.~L.\ 1963, Boletin de la Asociacion Argentina de Astronomia La Plata Argentina, 6, 41 
 \bibitem[Smol{\v c}i{\'c} et al.(2007)]{smo07} Smol{\v c}i{\'c}, V., Zucker, D.~B., Bell, E.~F., Coleman, M.~G., Rix, H.~W., Schinnerer, E., Ivezi{\'c}, {\v Z}., \& Kniazev, A.\ 2007, \aj, 134, 1901 
 \bibitem[Strigari et al.(2007)]{str07} Strigari, L.~E., Bullock, J.~S., Kaplinghat, M., Diemand, J., Kuhlen, M., \& Madau, P.\ 2007, \apj, 669, 676 
 \bibitem[Strigari et al.(2008)]{str08} Strigari, L.~E., Bullock, J.~S., Kaplinghat, M., Simon, J.~D., Geha, M., Willman, B., \& Walker, M.~G.\ 2008, \nat, 454, 1096 
 \bibitem[Tolstoy et al.(2004)]{tol04} Tolstoy, E., et al.\ 2004, \apjl, 617, L119 
 \bibitem[Tonini et al.(2006)]{ton06} Tonini, C., Lapi, A., \& Salucci, P.\ 2006, \apj, 649, 591 
 \bibitem[van den Bosch \& Swaters(2001)]{vdb01} van den Bosch, F.~C., \& Swaters, R.~A.\ 2001, \mnras, 325, 1017 
 \bibitem[Walker et al.(2007)]{wal07} Walker, M.~G., Mateo, M., Olszewski, E.~W., Gnedin, O.~Y., Wang, X., Sen, B., \& Woodroofe, M.\ 2007, \apjl, 667, L53 
 \bibitem[Walker et al.(2009a)]{wal09a} Walker, M.~G., Mateo, M., \& Olszewski, E.~W.\ 2009, \aj, 137, 3100 
 \bibitem[Walker et al.(2009b)]{wal09b} Walker, M.~G., Mateo, M., Olszewski, E.~W., Sen, B., \& Woodroofe, M.\ 2009, \aj, 137, 3109 
 \bibitem[Walker et al.(2009c)]{wal09c} Walker, M.~G., Mateo, M., Olszewski, E.~W., Pe{\~n}arrubia, J., Wyn Evans, N., \& Gilmore, G.\ 2009, \apj, 704, 1274 
 \bibitem[Wolf et al.(2009)]{wol09} Wolf, J, et al. 2009, \mnras, submitted (arXiv:0908.2995)


\end{thebibliography}
\end{document}